\documentclass[preprint]{aastex}
\slugcomment{To appear in ApJ-Letters}

\begin{document}
\title{A dynamical mechanism for establishing apsidal resonance}

\author{Renu Malhotra}
\affil{Department of Planetary Sciences, University of Arizona,
1629 E.~University Blvd., Tucson, AZ 85721,
renu@lpl.arizona.edu}

\begin{abstract}

We show that in a system of two planets initially in nearly circular orbits,
an impulse perturbation that imparts a finite eccentricity to one planet's
orbit causes the other planet's orbit to become eccentric as well,
and also naturally results in a libration of their relative apsidal longitudes
for a wide range of initial conditions.
We suggest that such a mechanism may explain orbital eccentricities and
apsidal resonance in some exo-planetary systems.  
The eccentricity impulse could be caused by the ejection of a planet from 
these systems, or by torques from a primordial gas disk.
The amplitude of secular variations provides an observational constraint 
on the dynamical history of such systems.

\end{abstract}


\newpage

\section{Introduction}

Orbital eccentricities in 
exo-planetary systems\footnote{http://www.exoplanets.org}
discovered thus far are often surprisingly large and have proven to be
a major puzzle in understanding these systems.
At least two systems with multiple planets in eccentric orbits
($\upsilon$ Andromedae and HD 83443) are suspected of exhibiting secular
apsidal resonance (Chiang, Tabachnik \& Tremaine 2001, Wu \& Goldreich 2002).
Apsidal resonance is the phenomenon of phase-locking of the apsidal longitudes
of two orbits, such that the two planets have a common average rate of apsidal
precession and the angular difference of their apsidal longitudes,
$\Delta\varpi=\varpi_1-\varpi_2$, librates around 0. 

We describe here a dynamical mechanism for establishing apsidal resonance
in a pair of planets that are initially on nearly circular orbits.
We use classical analysis from celestial mechanics to show that an impulse
perturbation that imparts an eccentricity to one of the orbits, excites the
other planet's eccentricity on a secular timescale and also results in the
libration of the relative apsidal longitude for a wide range of initial
conditions.  A plausible cause of an impulse perturbation is the ejection
of a planet from the system; torques from an exterior primordial disk may
also cause such an eccentricity impulse.

\section{Secular dynamics}

Consider a pair of planets in well-separated orbits, whose orbital period
ratio is not too close to a ratio of two small integers.  Under the assumption
that the relative inclination of their orbital planes is small, the secular
interactions in such a system preserve the orbital semimajor axes and induce
quasi-periodic variations in the other orbital parameters. For small
eccentricities and small relative inclinations, the eccentricity-apsidal line
variations are decoupled from the inclination-nodal line variations
(Murray \& Dermott 1999).  Under these conditions, the secular variations
of the eccentricities and apsidal longitudes are described by a set of first
order linear differential equations for the components of the so-called
eccentricity vectors, $(k_j(t),h_j(t))=e_j(\cos\varpi_j,\sin\varpi_j)$,
for each of the planets, where $e_j$ is the eccentricity and $\varpi_j$
is the apsidal longitude.  The general solution for two planets is a
superposition of two eigenmodes,
\begin{eqnarray}
h_j(t) &=& E_j^{(1)}\sin (g_1t+\beta_1)+ E_j^{(2)}\sin (g_2t+\beta_2), \cr
k_j(t) &=& E_j^{(1)}\cos (g_1t+\beta_1)+ E_j^{(2)}\cos (g_2t+\beta_2).
\end{eqnarray}
Let $m_\star$, $m_1$, $m_2$ be the masses of the star, the inner planet and
the outer planet, respectively; $n_1$, $n_2$ the orbital mean motions of the
inner and outer planet, respectively; $\alpha=a_1/a_2 <1$ the ratio of the
semimajor axes, $b_{s}^{(l)}(\alpha)$ a Laplace coefficient.  We also define
$\mu=m_1/m_2$, and $B=b_{3/2}^{(2)}/b_{3/2}^{(1)}$.
With these definitions, the mode frequencies
are 
\begin{equation}
g_{1,2} = {1\over8}\sqrt{\alpha}b_{3/2}^{(1)}(\alpha){m_2\over m_\star}n_2
\Big\{
1+\mu\sqrt\alpha\pm\Big[ (1-\mu\sqrt\alpha)^2+4\mu\sqrt\alpha B^2\Big]^{1/2}
\Big\},
\end{equation}
and the eigenvector amplitudes ${\bf E}^{(j)}$ are given, up to a
normalization factor, by 
\begin{equation}
{E^{(j)}_2\over E^{(j)}_1} = { 1-\mu\sqrt{\alpha}\mp
[(1-\mu\sqrt{\alpha})^2+4\mu\sqrt{\alpha}B^2]^{1/2}\over 2B}\equiv\rho_j,
\end{equation}
where the $`-'$ sign in $\mp$ corresponds to $j=1$, and the $`+'$ sign to $j=2$.
The phases $\beta_j$ and the normalization factors for the amplitudes of the
modes are determined by initial conditions.

We assume initially nearly circular orbits, as is expected of planet formation
in a dissipative protoplanetary disk.  And we consider an impulse
perturbation\footnote{In general, an impulse perturbation would result in not
only an eccentricity impulse but also impulse perturbation to other orbital
elements.  The secular dynamics described here remains valid provided we use
the post-impulse value of $a_2$ in evaluating $\alpha$ and $n_2$.} that
imparts a finite eccentricity, $e_{\rm 2f}$, to the outer planet's orbit,
at time $t=0$.  (The case of an impulse to the inner planet's orbit is solved
in a similar way.) Then, with initial conditions
$h_1(0)=h_{10},k_1(0)=k_{10},h_2(0)=0,k_2(0)=e_{\rm 2f}$,
we have
\begin{eqnarray}
E_1^{(1)} \cos\beta_1 &=& {\rho_2k_{10}-e_{\rm 2f}\over \rho_2-\rho_1},\qquad
E_1^{(1)} \sin\beta_1 = {\rho_2h_{10}\over \rho_2-\rho_1},\cr
E_1^{(2)} \cos\beta_2 &=& {\rho_1k_{10}-e_{\rm 2f}\over \rho_1-\rho_2},\qquad
E_1^{(2)} \sin\beta_2 = {\rho_1h_{10}\over \rho_1-\rho_2}.
\end{eqnarray}

\noindent
This solution admits either circulation or libration of the relative apsidal
longitude, $\Delta\varpi$.  Examples of three types of possible behavior are
shown in Fig.~1. In each of these cases, the cumulative probability density
function of $|\Delta\varpi|$ and of $e_1$ (i.e., the fraction of time that
$|\Delta\varpi|,e_1$ is less than some value) is also shown.

In the case of zero initial eccentricity, $e_{10}=0$ [Fig.~1(a--d)],
mode 1 and mode 2 appear with equal amplitude but opposite phase in the inner
planet's eccentricity vector. The time evolution of the planets' 
eccentricities and of $\Delta\varpi$ is given by
\begin{eqnarray}
e_1(t) &=& {\sqrt{2}e_{\rm 2f}\over|\rho_1-\rho_2|}
\Big[1-\cos(g_1-g_2)t\Big]^{1/2}, \cr
e_2(t) &=& {e_{\rm 2f}\over|\rho_1-\rho_2|}
\Big[\rho_1^2+\rho_2^2-2\rho_1\rho_2\cos(g_1-g_2)t\Big]^{1/2}, \cr
\tan\Delta\varpi(t)
&=& {\rho_2-\rho_1\over\rho_2+\rho_1} \tan[{1\over2}(g_1-g_2)t+{\pi\over2}].
\end{eqnarray}
The eccentricity of the inner planet grows from its initial
value of zero to a value comparable to $e_{\rm 2f}$ on the secular timescale
$\sim |g_1-g_2|^{-1}$; the eccentricities vary periodically (but not
sinusoidally) with a common frequency, $g_1-g_2$, and with opposite phases.
The inner planet's eccentricity has a maximum excursion to
$2e_{\rm 2f}/|\rho_1-\rho_2|$, while the eccentricity of the outer planet
varies between a minimum of $e_{\rm 2f}|\rho_1+\rho_2|/|\rho_1-\rho_2|$
and a maximum of $e_{\rm 2f}$.
The apsidal longitude difference is limited to $(-90^\circ,90^\circ)$;
it oscillates with the same frequency, $g_1-g_2$.
The singularity of the tangent function at $\pm{1\over2}\pi$
implies a discontinuity in the time evolution of $\Delta\varpi$.
This discontinuity is not unphysical (see Fig.~2 below):
as the eccentricity vector of the inner planet passes periodically through
the origin, its apsidal longitude changes discontinuously.
 
A small non-zero initial eccentricity, $e_{10}\ll e_{\rm 2f}$, results in
apsidal libration with an amplitude smaller than 90$^\circ$ for values
of $\varpi_{10}$ that are approximately within $\pm90^\circ$ of $\varpi_{20}$
[e.g., Fig.~1(e--h)], and apsidal circulation otherwise [e.g., Fig.~1(i--l)].
In all cases, the secular dynamics following an impulse perturbation to
$e_2$ yields a low probability of finding $|\Delta\varpi|>90^\circ$
[see Fig.~1(c,g,k)], and only $\sim\!1/3$ probability of finding $e_1$ less than
half its maximum [see Fig.~1(d,h,l)].
As we show below, a small non-zero $e_{10}$ results in apsidal resonance with
approximately 50\% probability.

It is illustrative to consider the limiting case $m_1\longrightarrow0$.
In this case, the orbital elements of the outer planet are constant:
$k_2(t)=e_{\rm 2f},h_2(t)=0$;
for the inner planet, we have the following solution:
\begin{eqnarray}
k_1(t) &=&+(k_{10}-e_{\rm 1f})\cos gt + h_{10}\sin gt + e_{\rm 1f}, \cr
h_1(t) &=&-(k_{10}-e_{\rm 1f})\sin gt + h_{10}\cos gt,
\end{eqnarray}
with
\begin{equation}
e_{\rm 1f}=B e_{\rm 2f}\qquad \hbox{and} \qquad
g= {1\over4}n_1{m_2\over m_\star}\alpha^2b_{3/2}^{(1)}(\alpha).
\end{equation}
This solution admits a simple geometrical interpretation as illustrated in
Fig.~2a: the inner planet's eccentricity 
is the vector sum of a `forced' component of magnitude $e_{\rm 1f}$
which is aligned with the outer planet's apsidal longitude (taken to be zero
here, without loss of generality), and a `free' component of magnitude
\begin{equation}
e_{\rm 1,free}=(e_{10}^2+e_{\rm 1f}^2-2e_{\rm 1f}k_{10})^{1/2}
\end{equation}
which rotates with frequency $g$.  Note that $e_{\rm 1f}$ is independent of
the mass of the outer planet, but the frequency $g$ is proportional to that
mass.

Apsidal libration occurs if $e_{\rm 1,free} \le e_{\rm 1f}$.
In the $(k_{10},h_{10})$ plane, this condition is satisfied in the area
of intersection of two circles, one of radius $e_{10}$ centered at the
origin, the other of radius $e_{\rm 1f}$ centered at $(e_{\rm 1f},0)$,
as shown in Fig.~2b.  Assuming random values of $\varpi_{10}$ and
$e_{10}\le\varepsilon$, the probability $P_{\rm lib}$ that the impulse on
the outer planet will result in apsidal libration is simply the intersection
area as a fraction of the area of the circle of radius $\varepsilon$.
This is readily evaluated:
\begin{equation}
P_{\rm lib}(e_{10}\le\varepsilon) =
{1\over\pi}\Big[
\cos^{-1}\Big({\varepsilon\over2e_{\rm 1f}}\Big)
+2\Big({e_{\rm 1f}\over \varepsilon}\Big)^2
 \sin^{-1}\Big({\varepsilon\over2e_{\rm 1f}}\Big)
-{e_{\rm 1f}\over \varepsilon}\sqrt{1-\Big({\varepsilon\over2e_{\rm 1f}}\Big)^2}
\Big].
\end{equation}
For $\varepsilon\ll e_{\rm 2f}$, we have $P_{\rm lib} =0.5$: nearly all 
values of $\varpi_{10}$ between $-90^\circ$ and $+90^\circ$ result in apsidal
resonance.  For $\varepsilon=e_{\rm 1f}$, the probability of apsidal resonance
drops only slightly, to $\sim\!0.4$.

For the general case, $m_1\ne0$, the probability of apsidal resonance
can be determined numerically.  It is easy to see from the general solution
(eq.~4) that $P_{\rm lib}$ has only a weak dependence on the mass ratio
$\mu=m_1/m_2$, and $P_{\rm lib}$ remains close to 0.5.

When apsidal libration occurs, the libration amplitude is a steep function
of $\varpi_{10}$ and $e_{10}/e_{\rm 2f}$.  This is illustrated in Fig.~3, for
one choice of parameters, $\mu=0.5$ and $\alpha=0.33$,
corresponding to the case of planets C and D of $\upsilon$ Andromedae.

\section{Discussion}

We have shown that in a two planet system, an impulse perturbation on the
eccentricity of the outer planet can excite the eccentricity of the inner
planet on a secular timescale, and will result in apsidal libration
with a $\sim\!50\%$ probability.
There are two significant characteristics of the secular dynamics following
an eccentricity impulse, whether or not it results in apsidal resonance:
(i) the inner planet's eccentricity has large amplitude variation,
and it drops to its initial small value periodically; 
(ii) only a small fraction of time is spent at large values of
$|\Delta\varpi|$ and small values of $e_1$ (see Fig.~1).
Therefore, 
observing small values of $|\Delta\varpi|$ and large values of $e_1$ is not
surprising if the system has suffered an impulse perturbation in its history
and the secular dynamics indicates large amplitude variations.

We note that in the opposite limit of a slow adiabatic perturbation which
increases one planet's eccentricity on a timescale much longer than the secular 
timescale, apsidal resonance will occur with nearly 100\% probability and
the resulting libration amplitude will be very small (for initially
circular orbits). 
It is rather remarkable that the secular dynamics of two planets admits a
high probability of apsidal resonance in both the impulse and the adiabatic
limit of eccentricity perturbation of one planet.
We defer detailed analysis to a future study, but we note here that
the apsidal libration amplitude provides an observational diagnostic:
large amplitudes favor the impulse mechanism whereas small amplitudes favor
the adiabatic mechanism.

One possible mechanism for the eccentricity impulse is {\it planet-planet 
scattering}.
Highly eccentric orbits can be produced by gravitational interactions of
planets initially on circular orbits that are close to the threshold of
orbital instability. 
This has been proposed as a possible explanation for the eccentric orbits
of exo-planets 
(Rasio \& Ford 1996, Weidenschilling \& Marzari 1996, Lin \& Ida 1997,
Ford et al.~2001, Marzari \& Weidenschilling 2002).
There are three possible outcomes of such interactions: a final stable
configuration different from initial, the merger of two planets, or the
ejection of one planet.
Numerical experiments thus far have explored only a small fraction of the
parameter space of two and three giant-planet systems.
In systems of two equal mass planets, comparable branching ratios for the
three possible outcomes are found; the two former outcomes do not result
in eccentricity excitation, but the third -- ejection of a planet -- does,
with typical eccentricity of the surviving planet in the range 0.4--0.8
(Ford et al.~2001).
One problematic aspect is that two planet systems must have initial orbits
fine-tuned close to the threshold of dynamical instability,
i.e. orbital separation close to $2\sqrt{3}$ times their mutual Hill radius
(Gladman 1993).
But this problem does not exist in systems of three or more planets, where
chaotic instabilities can occur over a wider range of orbital separations;
such systems can persist in quasi-stable orbital configurations for long
periods of time, possibly exceeding $10^8$ years, before becoming dynamically
unstable; then the most common outcome is the ejection of one planet
(Marzari \& Weidenschilling 2002).  Also, in the more realistic case of
unequal masses, the probability of planet ejection is likely to be enhanced
further.

We consider the ejection into an unbound orbit of
a third planet by gravitational scattering from the outer of the two planets
whose secular dynamics we have analyzed in section 2. 
(For comparable masses,
ejection by scattering from the outer of the two planets is more likely than
from the inner planet because the escape velocity is smaller at larger
distance from the star: $v_{\rm esc}\sim r^{-1/2}$; it is also likely to
produce the least direct perturbation to the inner members of the system.)
Assuming that the ejected planet leaves on a nearly parabolic orbit,
energy and angular momentum conservation yield the following estimate
for the final eccentricity of the bound planet:
\begin{equation}
1 - e_{\rm 2f}^2 \simeq (1+{m_3\over m_2} {a_2\over a_3})\Big[
1+ {m_3\over m_2}\sqrt{{m_\star+m_2\over m_\star+m_3}}
\Big(\sqrt{a_3\over a_2}-\sqrt{2q_3\over a_2}\Big) \Big]^2,
\end{equation}
where $m_3,a_3,q_3$ are the mass, initial semimajor axis and final periastron
distance of the ejected planet; $a_2$ is the initial semimajor axis of $m_2$,
and is related to its final semimajor axis by energy conservation:
$a_{\rm 2f} \approx m_2a_2a_3/(m_2a_3+m_3a_2)$.
Unfortunately, there is no additional dynamical constraint or simple argument
to constrain the final periastron distance of the ejected planet. So the
final eccentricity of the bound planet cannot be predicted analytically
and must be determined numerically.  Still, we can make the following 
estimate: since the periastron distance of the ejected planet is likely to be
not greatly in excess of the apoastron distance of the surviving planet,
then, for $q_3/a_2\simeq$ 1--1.5, the final eccentricity of the bound
planet will be in the range of 0.2--0.7 for $m_3/m_2\simeq$ 0.1--0.7.

Once a dynamical instability sets in, the subsequent evolution is highly
chaotic and unpredictable in detail.
A simplified description, based upon the presently available numerical
simulations, is as follows:
at first, close encounters between $m_2$ and $m_3$ lead to a rapid increase
in their orbital eccentricities, over $\sim\!10-10^2$ orbital periods;
this is followed by a longer period, $\gtrsim 10^4$ orbits, during which many
weak distant encounters gradually and stochastically increase the orbital
period and apoastron of $m_3$, the planet-to-be-ejected.
In the present context, the initial eccentricity excitation of the surviving
planet, $m_2$, satisfies the impulse approximation as the timescale of the
first stage is much shorter than the secular timescale (which is
of order $(m_{1,2}/m_\star)^{-1}$ times the orbital period,
or $\sim\!10^3$ orbital periods for jovian mass planets).
Thus, the secular dynamics of the two surviving planets, $m_1$ and $m_2$,
would be as described in Section 2.
The weak distant encounters that eventually lead to the ejection of the
third planet would produce small perturbations to the secular solution,
as would the gravitational interactions between $m_1$ and $m_3$.
(Neglecting the latter is justified, in the lowest order, as the
$m_1$,$m_3$ separation is much greater than the $m_2$,$m_3$ separation.)
The effects of these perturbations on the secular dynamics presented above
will be considered in a future study.
\medskip

\noindent{\it The case of $\upsilon$ Andromedae}\quad
The analysis given in the previous section can be used to estimate the
pre-impulse orbital eccentricity of at least one of the planets.
A good candidate amongst the known exo-planetary systems to apply this theory
is the $\upsilon$ Andromedae system, which represents perhaps the most
(dynamically) constrained of the known exo-planetary systems.
A recent orbital solution for this system (as quoted in Chiang et al.~2001)
gives present eccentricities of planets C and D of 0.25 and 0.34, 
respectively, and yields secular dynamics for these two planets similar to 
that shown in Fig.~1(e,f), with apsidal libration amplitude of
25$^\circ$--35$^\circ$ (for relative orbital inclinations $\la 20^\circ$).
An independent orbital solution obtained previously by Stepinski et al.~2000
(from an older and therefore smaller set of observational data) has orbital 
parameters within $\sim\!2\sigma$ uncertainties of the more recent solution;
it yields secular dynamics similar to that shown in Fig.~1(a,b), with apsidal
libration amplitude of 80$^\circ$--90$^\circ$.
The libration period for this system is $\sim7\times10^3$ yr.
Now, if we allow that planet D suffered an eccentricity impulse,
$e_D\approx0.35$, in its history, we can estimate that the initial
eccentricity of planet C could have been as small as $\sim\!0.07$ if
the apsidal libration amplitude is 25--35 deg, but even smaller,
0--0.01, if the amplitude is 80--90$^\circ$.

After submitting this paper, we became aware of a recent preprint by 
Chiang \& Murray (2002) who propose an adiabatic eccentricity perturbation
to explain the apsidal resonance in the $\upsilon$ Andromedae planetary system.
They invoke torques from an exterior massive primordial disk to provide the
adiabatic eccentricity excitation to the outermost planet D.
However the adiabaticity of the perturbation is owed to a specific choice
of disk parameters, which are not well constrained; a different, and perhaps
equally plausible, choice of disk parameters may instead provide an impulse
perturbation (E.~Chiang 2002, personal communication);
then the secular dynamics described here would apply.

As mentioned previously, the apsidal libration amplitude provides an
observational diagnostic: large amplitudes favor the impulse mechanism whereas
small amplitudes favor the adiabatic mechanism.
In the case of $\upsilon$ Andromedae, there remain significant uncertainties
in the orbital parameters, and at present it appears difficult to discriminate
between the adiabatic and the impulse mechanisms.
The 25--35 deg apsidal libration amplitude found by Chiang et al.~(2001)
is neither clearly small nor clearly large; in both the adiabatic mechanism
and the impulse mechanism, an initial eccentricity near 0.06--0.07 of planet C 
can account for this libration amplitude.
On the other hand, the large libration amplitude of 80--90 deg found by 
Stepinski et al.~(2000) would be explained by initial $e_C\lesssim 0.01$ in 
the impulse limit, but cannot be explained naturally in the adiabatic limit.
We urge further observations and analysis to improve the accuracy and fidelity 
of orbital solutions of exo-planetary systems; this would help to constrain
their dynamical history.

\acknowledgements
We acknowledge comments from E.~Chiang, discussions with F.~Rasio,
and research support from NASA.

\begin{figure}
\plotone{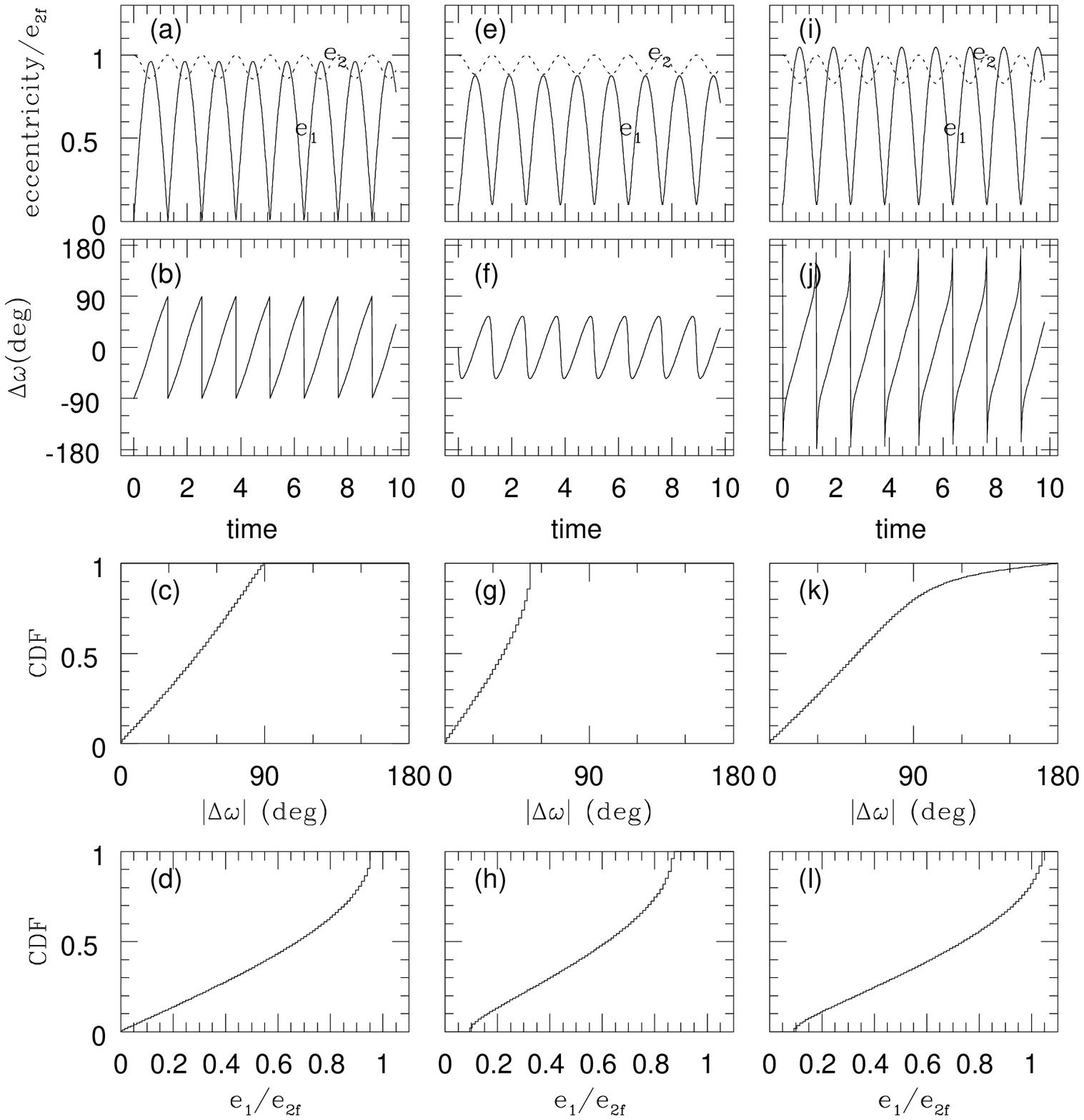}
\caption{Secular dynamics for $\mu=0.5, \alpha=0.33$ and the following initial
conditions: (a)--(c) $e_{10}=0, \varpi_{10}=0$;
(d)--(f) $e_{10}=0.1 e_{\rm 2f}, \varpi_{10}=0$;
(g)--(i) $e_{10}=0.1 e_{\rm 2f}, \varpi_{10}=\pi$.
The upper two panels show the time variation of the eccentricities $e_1,e_2$,
and the difference of apsidal longitudes, $\Delta\varpi=\varpi_1-\varpi_2$;
the unit of time is $2\pi/g_1$.  The lower two panels plot the cumulative
probability density function of $|\Delta\varpi|$ and of $e_1$.}
\end{figure}

\vfill\eject

\begin{figure}
\plotone{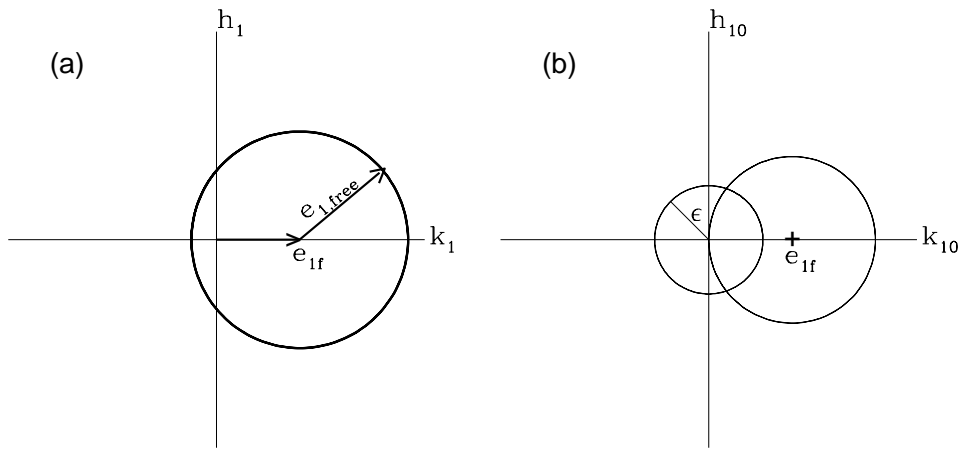}
\caption{For $m_1=0$, the secular variation of $m_1$'s eccentricity vector
is the vector sum of a forced component of magnitude $e_{\rm 1f}$ along the
direction of $m_2$'s apsidal longitude and a rotating free component of
magnitude $e_{\rm 1,free}$ (a).  For initial $e_{10}<\varepsilon$, apsidal
resonance occurs for initial conditions $(k_{10},h_{10})$ in the area of
intersection of the two circles shown in (b).}
\end{figure}
\vfill\eject

\begin{figure}
\plotone{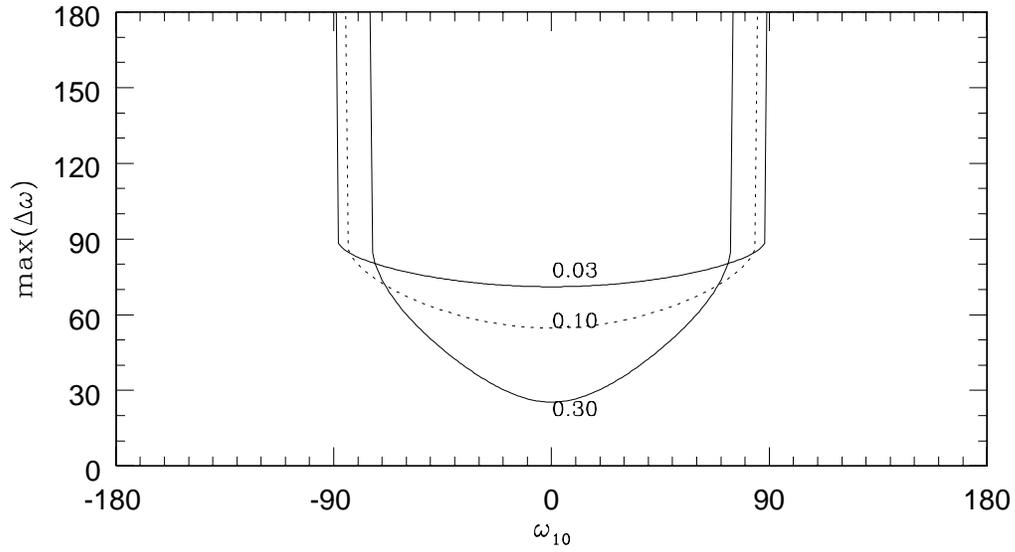}
\caption{Apsidal libration amplitude as a function of initial $\varpi_{10}$
for $\mu=0.5, \alpha=0.33$, and three different values of
$e_{10}/e_{\rm 2f}=0.03,0.1,0.3$.}
\end{figure}

\end{document}